\documentclass[
 reprint,
 amsmath,amssymb,
 aps,
 citeautoscript
]{revtex4-1}
\usepackage{graphicx,color}
\usepackage{float}
\usepackage[caption = false]{subfig}
\usepackage{epsfig}
\usepackage{graphicx}
\usepackage{mathrsfs}
\usepackage{color}
\usepackage{bm}
\usepackage{amsmath,amssymb,amsthm,amscd}
\usepackage{bbold}
\usepackage{mdwlist}

 \newcommand{\braket}[2]{\langle{#1}| {#2}\rangle}
 \newcommand{\ketbra}[2]{\vert {#1} \rangle \langle{#2}\vert}

\newcommand{\Tr}{\operatorname{Tr}}

\allowdisplaybreaks

\begin{document}

\title{Bounding the Classical Capacity of Multilevel Damping Quantum Channels}

\author{Chiara Macchiavello}
\email{chiara.macchiavello@unipv.it}
\affiliation{Quit group, Dipartimento di Fisica, 
Universit\`a di Pavia, via A. Bassi 6, 
 I-27100 Pavia, Italy}
\affiliation{Istituto Nazionale di Fisica Nucleare, Gruppo IV -
  Sezione di Pavia, via A. Bassi 6,
  I-27100 Pavia, Italy}
\affiliation{Istituto Nazionale di Ottica - CNR, Largo E. Fermi 6, I-50125, Firenze, Italy}

\author{Massimiliano F. Sacchi}
\email{msacchi@unipv.it}
\affiliation{Istituto di Fotonica e Nanotecnologie - CNR, Piazza Leonardo
  da Vinci 32, I-20133, Milano, Italy}
\affiliation{Quit group, Dipartimento di Fisica, 
Universit\`a di Pavia, via A. Bassi 6, 
I-27100 Pavia, Italy}

\author{Tito Sacchi}
\email{tito.sakki@gmail.com}
\affiliation{Istituto Superiore ``Taramelli-Foscolo'', via L. 
  Mascheroni 51, I-27100 Pavia, Italy}

\date{\today}

\begin{abstract} 
A recent method to certify the classical capacity of quantum
communication channels is applied for general damping channels in
finite dimension. The method compares the mutual information
obtained by coding on the computational and a Fourier basis, which can
be obtained by just two local measurement settings and classical
optimization. The results for large representative classes of
different damping structures are presented.
\end{abstract}

\maketitle
\section{Introduction}
The complete characterization of quantum communication channels by
quantum process tomography \cite{nielsen97,mohseni,irene} becomes
demanding in terms of state preparation or measurement settings
for increasing dimension $d$ of the system Hilbert space since it
scales as $d^4$.  Actually, growing interest has been shown recently
for quantum communication protocols based on larger alphabets, beyond
the binary case with $d=2$, since they can offer advantages with
respect to the two-dimensional case, from higher information capacity
to increased resilience to noise
\cite{info1,info2,info3,info4}. Several physical systems allow
encoding of higher dimensional quantum information, e.g. Rydberg atoms
\cite{ryd}, cold atomic ensembles \cite{cold1,cold2}, polar molecules
\cite{pol}, trapped ions \cite{tra}, NMR systems \cite{nmr0}, photon
temporal modes \cite{christine} and discretized degrees of freedom of
photons \cite{zeil}.  Hence, as the size of quantum devices continues
to grow, the development of scalable methods to characterise and
diagnose noise is becoming an increasingly important problem.

In some situations one is experimentally interested in characterizing
only specific features of an unknown quantum channel. Then, less demanding
procedures can be adopted with respect to complete process tomography, as for example
in the case of detection of
entanglement-breaking properties \cite{qchanndet,qchanndet2} or
non-Markovianity \cite{nomadet} of quantum channels, or for detection
of lower bounds to the quantum capacity \cite{ms16,ms-corr,exp,mixed}.
In fact, some properties by themselves are not directly accessible
experimentally, as for example the ultimate classical capacity of quantum channels, which
generally requires a regularisation procedure over an infinite number
of channel uses \cite{NC00,hol0,sw,hol}. Moreover, by adopting quantum process
tomography to reconstruct just a single use of the channel, we notice that
the evaluation of the classical capacity remains a theoretically hard
task, even numerically \cite{bs,eff,eff2,eff3,eff4,eff5}.

It is therefore very useful to develop efficient means to establish whether a communication 
channel can be profitably employed for information transmission
when the kind of noise affecting the channel is not known.
For the
purpose of detecting lower bounds to the classical capacity a versatile
and proficient procedure has been recently presented in
Ref. \cite{ms19}.  The method allows to experimentally detect useful lower bounds to
the classical capacity by means of few local measurements, even for
high-dimensional systems. The core of the procedure is to efficiently
measure a number of probability transition matrices for suitable input
states and matched output projective measurements, and then to
evaluate the pertaining mutual information for each measurement
setting. This is achieved by finding theoretically or numerically the
optimal prior distribution for each single-letter encoding. Hence, a
lower bound to the Holevo capacity and then a certification of minimum
reliable transmission capacity is achieved.

In this paper we apply the above method to detect lower bounds to the
classical capacity of general damping channels in dimension $d>2$.
The form of channels we consider has been previously investigated in 
the context of quantum error correcting codes \cite{debbie,ieee}.
We will compare the mutual information achieved by coding on the
computational and a Fourier basis, which can be obtained by just two
local measurement settings and classical optimization.
We present the results for large representative classes of different damping
structures for high-dimensional quantum systems.

\section{The general method}
We briefly review the method proposed in Ref. \cite{ms19}.
The classical capacity $C$ of a noisy quantum channel
${\cal E}$ quantifies the maximum number of bits per channel use that
can be reliably transmitted. It is
defined \cite{hol0,sw,hol} by the regularized expression $C=\lim
_{n\rightarrow \infty} \chi ({\cal E}^{\otimes n})/n$, in terms of the
Holevo capacity
\begin{eqnarray}
\chi (\Phi) = \max _{\{p_i , \rho _i \}}\{S[\Phi (\textstyle \sum _i p_i \rho
  _i)]-\textstyle 
\sum
_i p_i S[\Phi (\rho _i)]\}
\;,
\end{eqnarray}
where the maximum is computed over all possible ensembles of quantum states,
and $S(\rho )=-\Tr[\rho \log \rho ]$ denotes the von Neumann entropy
(we use logarithm to base $2$). The Holevo capacity $\chi ({\cal
  E})\equiv C_1$ is a lower bound for the channel capacity, and
corresponds to the maximum information when only product states are
sent through the uses of the channel, whereas joint (entangled)
measurements are allowed at the output. Then, clearly, the Holevo
capacity is also an upper bound for any expression of the mutual
information \cite{mutu,mutu2,mutu3}
\begin{eqnarray}
I(X;Y)=\sum _{x,y}p_x p(y|x) \log \frac {p(y|x)}{\sum _{x'} p_{x'} p(y
  |x')}\;,
\end{eqnarray}
where the transition matrix $p(y|x)$ corresponds to the
conditional probability for outcome $y$ in an arbitrary measurement 
at the output for a single use of the channel with input $\rho _x$,
and $p_x$ denotes an arbitrary prior probability, which describes the
distribution of the encoded alphabet on the quantum states $\{ \rho _x
\}$.
 
In order to detect a lower bound to the classical capacity when
the number of measurement settings is smaller than the one needed for
complete process tomography, the following strategy can be adopted. Prepare a
bipartite maximally entangled state $|\phi^+\rangle =\frac {1}{\sqrt
  d} \sum_{k=0}^{d-1} |k \rangle |k \rangle $ of a system and an
ancilla $A$ with the same dimension $d$; 
send $|\phi ^+ \rangle $ through the unknown channel ${\cal
  E}\otimes{\cal I} _A$, where ${\cal E}$ acts on
the system alone; finally, measure locally a number of observables of
the form $X_i\otimes X_i ^{\tau}$, where $\tau $ denotes the transposition
w.r.t. to the fixed basis defined by $|\phi ^+ \rangle $.

By denoting the $d$ eigenvectors of $X_i$ as $\{|\phi ^{(i)}_n \rangle
\}$ and using the identity \cite{pla}
\begin{equation}
\Tr [(A\otimes B^\tau )({\cal E}\otimes{\cal I} _R)
|\phi ^+ \rangle \langle \phi ^+ |]=\frac 1d   
\Tr[A {\cal E}(B)]\;,
\end{equation} 
the measurement protocol allows to reconstruct the set of
conditional probabilities $p^{(i)}(m|n) = \braket {\phi ^{(i)}_m} {
  {\cal E} (\ketbra {\phi ^{(i)}_n} {\phi ^{(i)}_n}) | {\phi ^{(i)}_m}
}$.  We can then write the optimal mutual information for the
encoding-decoding scheme by the observable $X_i$ as
\begin{eqnarray}
I^{(i)}=\max _ {\{p_n^{(i)} \}} \sum _{n,m}p_n^{(i)} 
p^{(i)}(m|n)  
\log \frac{p^{(i)}(m|n)  }{\sum _l 
p_l^{(i)} p^{(i)}(m|l) }
\;.\label{ii}
\end{eqnarray}
Then, the following chain of inequalities holds
\begin{eqnarray}
C 
\geq C_1 \geq C_{DET} \equiv \max _i \{I^{(i)}\}
\;,
\end{eqnarray}
where $C_{DET}$ is the experimentally accessible bound to the
classical capacity, which depends on the chosen set of measured
observables labeled by $i$.

Notice that such a detection method based on the
measurements of the local operators does not necessarily require the
use of an entangled bipartite state at the input. Actually, each
conditional probability $p^{(i)}(m|n)$ can be 
equivalently obtained  by
testing only the system, i.e. preparing it in each of the eigenstates of
$X_i$, and measuring $X_i$ at the output of
the channel. 

The maximisation over the set of prior probabilities $\{p_n ^{(i)} \}$
in Eq. (\ref{ii}) for each $i$ can be achieved by means of the
Blahut-Arimoto recursive algorithm \cite{bga1,bga2,bga3}, given by
\begin{eqnarray}
&&g_n ^{(i)}[r]
=\exp \left (\sum _m p^{(i)}(m|n)  \log \frac{p^{(i)}(m|n)  }{\sum _l 
p_l^{(i)}[r] p^{(i)}(m|l) } \right ) \,;
\nonumber \\& & 
p_n ^{(i)}[r+1]=p_n ^{(i)}[r] \frac {g_n ^{(i)}[r]}
{\sum _l p_l ^{(i)}[r] g_l ^{(i)}[r]} 
\;.\label{ari}
\end{eqnarray}
Starting from an arbitrary prior probability distribution $\{p_n ^{(i)} [0]\}$, 
this guarantees convergence to an optimal prior $\{ \bar p_n ^{(i)}\}$, thus providing the
value of $I^{(i)}$ for each $i$ with the desired accuracy.  A minor
modification of the recursive algorithm (\ref{ari}) can also
accommodate possible constraints, e.g. the allowed maximum energy in
lossy bosonic channels \cite{loss}.

\par We remind that for some special forms of transition matrices
$p^{(i)}(m |n)$ there is no need for numerical maximisation, since the
optimal prior is theoretically known. This is the case of a
conditional probability $p^{(i)}(m|n)$ corresponding to a symmetric
channel \cite{CT}, where every column $p^{(i)}(\cdot |n)$ [and row
$p^{(i)}(m |\cdot )$] is a permutation of each other. In fact, in such
a case the optimal prior is the uniform $\bar p^{(i)}_n=1/d$, and the
pertaining mutual information is given by $I^{(i)}=\log d -
H[p^{(i)}(\cdot | n)]$, where $H(\{x _j\})=-\sum _j x_j \log x_j$
denotes the Shannon entropy and therefore $H[p^{(i)}(\cdot | n)]$ is
the Shannon entropy of an arbitrary column (since all columns have the
same entropy).

\section{Multilevel damping channels}
We apply the general method summarized in the previous Section
to quantum channels of the Kraus form 
\begin{eqnarray}
{\cal E }(\rho )= \sum _{k=0}^{d-1} A_k \rho A^\dag _k\;,\label{er}
\end{eqnarray}
with
\begin{eqnarray}
A_k = \sum_ {r=k}^{d-1} c_{r-k,r} |r-k \rangle
\langle r |\;.\label{ak}
\end{eqnarray}
The trace preserving condition $\sum _{k=0}^{d-1}A^\dag _k A_k =I$
corresponds to the following constraints
\begin{eqnarray}
\sum_{k=0}^r |c _{r-k,r}|^2=1 \qquad {\hbox{for all }} r\,.
  \;
\end{eqnarray}
The above channels represent a generalization of damping channels for
$d$-dimensional quantum systems, where each level can populate only
its lower-lying levels, and no reverse transition can occur. This form
of channel can thus accommodate different decay processes, from
multilevel atoms to dissipative bosonic systems (see for example \cite{zubairy,puri}).  
The customary amplitude damping channel for qubits is recovered for $d=2$ and
$c_{0,0}=1$, $c_{0,1}=\sqrt{\gamma}$ and $c_{1,1}=\sqrt{1-\gamma}$. 
\par Typically, for a fixed value of $r$ each column vector $c_{r-k,r}$ will depend
on a set of damping parameters such that in a suitable limit for all
values of $r$ one has $c_{r-k,r}=\delta _{k,0}$  (or $c_{r-k,r}=\delta _{k,0} e^{i \psi
  _r}$). In this way, for such a limit
one obtains the noiseless identity map $\cal E(\rho )=\rho$ (or
noiseless unitary map ${\cal E}(\rho )=U \rho U^\dag $ where $U=\sum
_{r=0}^{d-1}e^{i\psi _r} |r \rangle \langle r|$) \cite{nota}. Notice
also that if the number of allowed jumps in the level structure is
limited to $S$, one will always have $c_{r-k,r}=0$ for $k>S$.
\par We consider the simplest case where only two projective measurements are used
to bound the classical capacity, namely the two mutually unbiased bases 
\begin{eqnarray}
&&B  =\left \{ |n \rangle, \  n\in [0,d-1] \right \} \,,\\& & 
\tilde B =\left \{| \tilde n \rangle  = \frac {1}{\sqrt d}
(\textstyle \sum _{j=0}^{d-1} \omega ^{nj} |j \rangle ),\ 
n\in [0,d-1]\right \}\,,\label{bunodue}
\end{eqnarray}
with $\omega =e^{2\pi i/d}$. 
The corresponding transition matrices for ``direct coding'' (with basis
$B$) and ``Fourier coding'' (with basis $\tilde B$) are given by
\begin{eqnarray}
 && Q(m|n)= \langle m | {\cal E }(|n \rangle \langle n |) |m \rangle \,,
  \\& &
  \tilde Q (m |n)= \langle \tilde m | {\cal E} (| \tilde n \rangle
  \langle \tilde n |) |
\tilde m \rangle \,,  
\end{eqnarray}
respectively. As we have seen, each of these transition matrices can
be experimentally reconstructed by preparing a bipartite maximally
entangled state and performing two separable measurements at the
output of the channel (which acts just on one of the two systems), or
equivalently by testing separately the ensemble of basis states with
the respective measurement at the output. The detected lower bound
$C_{DET}$ to the classical capacity of the channel then corresponds to
the larger value between $I^{(B)}$ and $I^{(\tilde B)}$, which are
obtained by Eq. (\ref{ii}).  \par The present study is inspired by a
specific case of damping channel for qutrits studied in
Ref. \cite{ms19}, where a transition between two different encodings
has been observed as a function of the damping parameters.  For increasing
dimension, the number of parameters characterizing the channel
increases and the solution can become quite intricate. We remind that
for the customary qubit damping channel no transition occurs, and the
Fourier basis always outperforms the computational basis \cite{ms19}.
\par From Eqs. (\ref{er}) and (\ref{ak}) one easily obtains the
identity
\begin{eqnarray}
  \langle m | {\cal E} (|n \rangle \langle l |) |s \rangle =
c_{m,n} c^*_{s,l} \delta _{l-s,n-m}
  \;.
\end{eqnarray}
Then, one has
\begin{eqnarray}
Q(m|n)=|c_{m,n}|^2 \;,
\end{eqnarray}
and
\begin{eqnarray}
  \!\!\!\!\!\!
  \tilde Q (m|n)= \frac {1}{d^2}\sum _{l=0}^{d-1}\sum _{s=0}^l \!\sum
_{t=0}^{d-1-l+s} \!\!\! c_{s,l}c^*_{t,l-s+t}\, 
\omega ^{(t-s)(m-n)}\!.  
\end{eqnarray}
Notice that $\tilde Q(m|n)$
just depends on $(m-n)\!\!\!\mod d$ and hence
it has the form of a conditional probability pertaining to a
symmetric channel. As noticed in the previous Section, in this
case the optimal prior distribution achieving the maximisation in Eq. (\ref{ii})
is always the uniform one,
and the corresponding mutual information is given by $I ^{(\tilde B)}=\log d -
H[\tilde Q (\cdot | n)] $.

On the other hand, the optimal prior distribution $\{\bar p_n \}$ for the
direct-basis coding can be obtained by the algorithm (\ref{ari}). In
this case, as a global measure of the non-uniformity of 
$\{\bar p_n \}$ one can consider its Shannon entropy $H(\{\bar p_n\})$. Clearly,
one has $0\leq H(\{ \bar p_n \})\leq \log d$.

Notice that for the direct basis, all channels considered here are
such that the output states commute with each other. Since the
Holevo bound to the accessible information is saturated for sets of
commuting states \cite{petz}, the detected
capacity for the direct basis coincides with the Holevo quantity,
namely 
\begin{eqnarray}
I^{(B)}&&=\chi_{B} \equiv S[{\cal E}(\textstyle \sum  _n \bar p_n |n \rangle \langle n
|)]-\textstyle \sum _n \bar p_n S[{\cal
    E}(|n \rangle \langle n|)]\nonumber \\& &
  = H[\textstyle \sum _n \bar p_n Q(\cdot|n )] -\textstyle \sum _n
  \bar p_n
H[Q(\cdot |n)]
\;, \label{holdir}
\end{eqnarray}
where $\{ \bar p_n \}$ denotes the optimal prior obtained by the
Blahut-Arimoto algorithm. 
On the other hand, the detected capacity for the Fourier basis
$I^{(\tilde B)}$ will be
bounded by the Holevo quantity, namely 
\begin{eqnarray}
I^{(\tilde B)}&& \leq \chi_{\tilde B} \equiv S[{\cal E}(\textstyle
    \sum _n \frac 1d |\tilde n \rangle \langle \tilde n |)]-\textstyle
  \sum _n \frac 1d S[{\cal E}(|\tilde n \rangle \langle \tilde
    n|)]\nonumber \\& & = S\left (\frac 1d \textstyle \sum _n \tilde
  \rho _n \right ) - \frac 1d \textstyle \sum _n S(\tilde \rho _n)
  \;,\label{holtil}
\end{eqnarray}
where
\begin{eqnarray}
  \tilde \rho _n &&= {\cal E}(|\tilde n \rangle \langle \tilde
  n|)
  \\& & =\frac 1d 
  \sum _{m,s=0}^{d-1}|m \rangle \langle s|
\sum_{t=m}^{d-1-s+m} c_{m,t}c^*_{s,s+t-m}\,\omega ^{n(m-s)}
  \;.\nonumber
\end{eqnarray}
Notice that
\begin{eqnarray}
\textstyle \sum_{n=0}^{d-1}\tilde \rho _n =\textstyle \sum_{m=0}^{d-1} |m \rangle \langle m
| (\textstyle \sum _{t=m}^{d-1} |c_{m,t}|^2)
  \;,
\end{eqnarray}
and hence the first term in Eq. (\ref{holtil}) is just given by 
\begin{eqnarray}
  S\left (\frac 1d \textstyle \sum _n \tilde \rho _n \right )
  = H (\{ w_m \})  \;,
\end{eqnarray}
with
\begin{eqnarray}
w_m = \frac 1d \textstyle \sum _{t=m}^{d-1} |c_{m,t} |^2
\;.
\end{eqnarray}

Clearly, the maximum between the two Holevo quantities (\ref{holdir})
and (\ref{holtil}) provides a better lower bound than $C_{DET}$ to the
ultimate classical capacity, but for an unknown quantum channel their
evaluation needs complete process tomography. We will
consider the values of $\chi _B$ and $\chi _{\tilde B}$ in order to
compare the results of the proposed method with a theoretical bound,
since the damping channels in dimension $d>2$ are theoretically poorly
studied and largely unexplored.

In the following we present numerical results for 
different multilevel damping channels, which explore many illustrative
scenarios. For simplicity, we will fix the matrix elements of
$c_{m,n}$ as real. This restriction is always irrelevant as regards the
direct basis. For $\arg c_{m,n}=f(n-m)$, this also holds for the
Fourier basis.

\subsection{Bosonic dissipation}
For a bosonic system with energy dissipation the damping structure is 
typically governed by Binomial distributions, namely 
\begin{eqnarray}
Q(m|n)= \binom{n}{m}\gamma _n ^{n-m}  (1- \gamma _n)^{m}  
\;.
\end{eqnarray}
In principle, notice that each level can be characterized by its own
damping parameter $\gamma _n \in [0,1]$. For this model of noise the
classical capacity is known \cite{cc} for infinite dimension with
mean-energy constraint and $\gamma _n =\gamma $ for all values of $n$. The mean
and variance of these distributions are given by
\begin{eqnarray}
  &&\mu _n = n(1-\gamma _n)\,, \\& &
  \sigma ^2 _n = n \gamma_n (1-\gamma _n)\,.
\end{eqnarray}
In Figs. 1-3 we present the results of the optimization for the
simplest case of $\gamma _n =\gamma $ for all values of $n$.  We notice that for
all values of $\gamma $ and any dimension $d$ the detected classical
capacity $C_{DET}$ depicted in Fig. 1 is achieved by the Fourier
encoding $\tilde B$. In Fig. 2, for $d=8$, we also report the best 
theoretical lower bound given by the Holevo quantity $\chi _{\tilde
  B}$ of Eq. (\ref{holtil}) and the looser bound obtained by the
direct basis $B$.
\begin{figure}[htb]
  \includegraphics[scale=0.46]{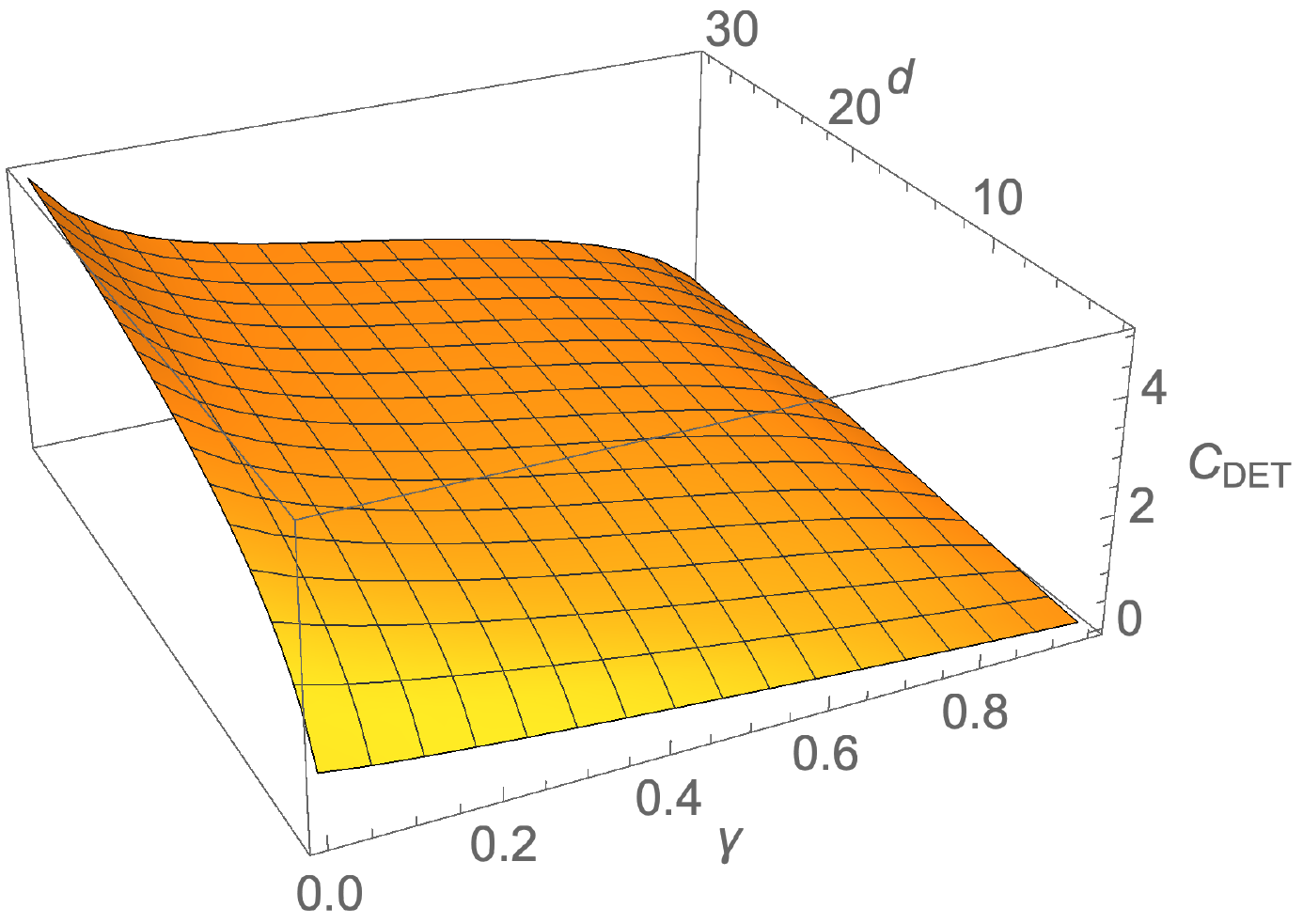}
  \caption{Detected classical capacity $C_{DET}$ (achieved by
    the Fourier basis $\tilde B$) for a bosonic dissipation channel
    vs dimension $d$ and damping parameters $\gamma_n=\gamma $.}
\end{figure}
\par\noindent In Fig. 3 we plot the rescaled difference
\begin{eqnarray}
\Delta =\frac{\chi_{\tilde B} -C_{DET}}{\log d}
  \;,\label{dd}
\end{eqnarray}
in order to compare the detected capacity with the
Holevo quantity $\chi _{\tilde B}$. 
\begin{figure}[htb]
  \includegraphics[scale=0.46]{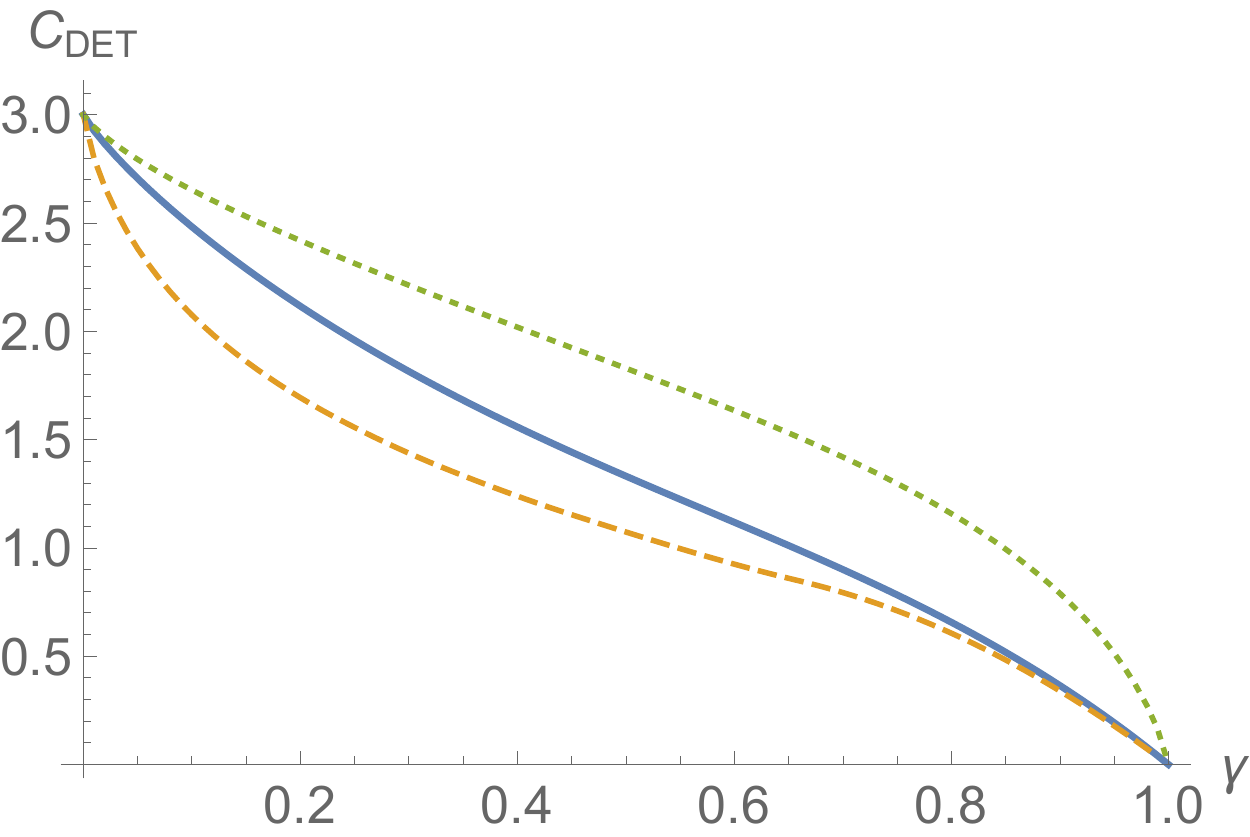}
  \caption{Detected classical capacity $C_{DET}$ for a bosonic
    dissipative channel vs damping parameters $\gamma _n = \gamma $ for
    $d=8$ (solid line, achieved by the Fourier basis $\tilde B$). The
    looser bound in dashed line corresponds to the direct basis
    $B$. The dotted line represents the theoretical lower bound given
  by the Holevo quantity $\chi _{\tilde B}$ of Eq. (\ref{holtil}).}
  \includegraphics[scale=0.52]{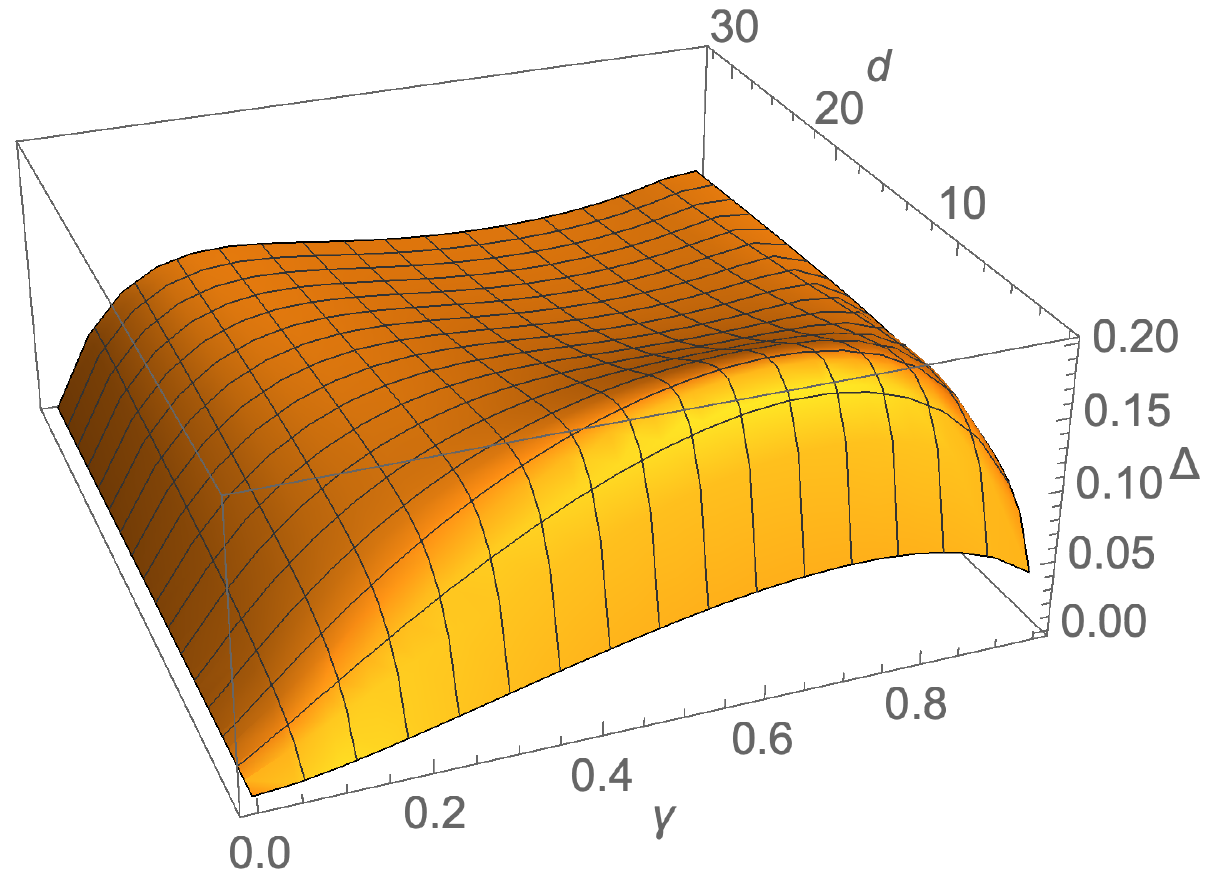}
  \caption{Rescaled difference $\Delta $ between the theoretical Holevo
    quantity $\chi _{\tilde B}$  and the detected classical capacity
    $C_{DET}$ for a bosonic dissipation channel
    vs dimension $d$ and damping parameters $\gamma_n=\gamma $.}
\end{figure}

\subsection{Hypergeometric channel}
We consider here a damping channel with decay structure characterized
by hypergeometric distributions, namely
\begin{eqnarray}
  Q(m|n)=
\frac {\binom{M}{m} \binom{L- M}{n - m}}
  {\binom{L}{n}} 
  \;.
\end{eqnarray}
with integer $M$ and $L$, with $0\leq M \leq L$ (in principle, both
$M$ and $L$ could vary for different values of $n$). This distribution
is related to the probability of $m$ successes in $n$ draws without
replacement from finite samples of $L$ elements, differently from the
binomial distribution where each draw is followed by a
replacement. The correspondence with the customary binomial
distribution is obtained for $M/L = 1-\gamma $. In fact, the mean and
variance are given by \cite{rice}
\begin{eqnarray}
  &&\mu = n \frac M L \;,\\& &
 \sigma ^2 = n \frac ML \left (1- \frac M L \right ) \frac {L-n}{L-1}
  \;.
\end{eqnarray}
Notice that the variance is shrunk by the factor $\frac {L-n}{L-1}$
with respect to the binomial distribution.  For $M,L \rightarrow
\infty$ with $M/L =p$ one recovers the binomial distribution.  We also
observe that the support of the distribution is given by $m \in \{
\max (0,n+M-L),\min (n,M)\}$. For $M/L=1$ the channel is lossless,
i.e. $c_{m,n}=\delta _{m,n}$. 
\par In Fig. 4 we report the result of the optimization for $d=8$ and
$L=12$ vs $M$. Differently from the case of bosonic
dissipation, one can observe a transition from the Fourier basis
$\tilde B$ to the direct basis $B$ in providing the best detected
capacity, for increasing value of the damping parameter
(i.e. for decreasing value of $M$ for fixed $L$). 

\begin{figure}[htb]
  \includegraphics[scale=0.46]{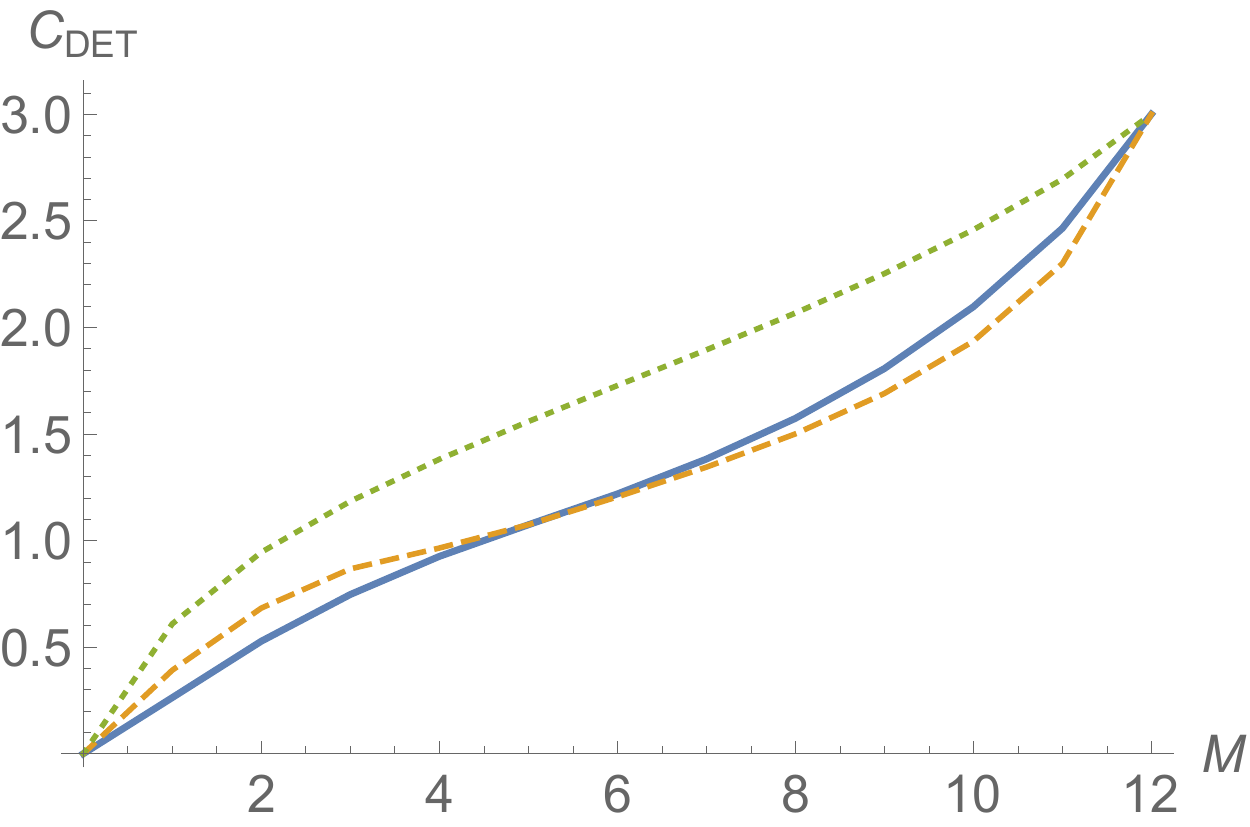}
  \caption{Detected classical capacity $C_{DET}$ for a damping channel
    with hypergeometric decay vs parameter $M$, with $0\leq M \leq
    L=12$, and dimension $d=8$. The detected capacity is achieved by
    the Fourier Basis (solid line) for $M\geq 6$, and by the direct
    basis $B$ (dashed line) for $M\leq 5$. The dotted line represents
    the theoretical lower bound given by the Holevo quantity $\chi
    _{\tilde B}$ of Eq. (\ref{holtil}).}
%
  \includegraphics[scale=0.46]{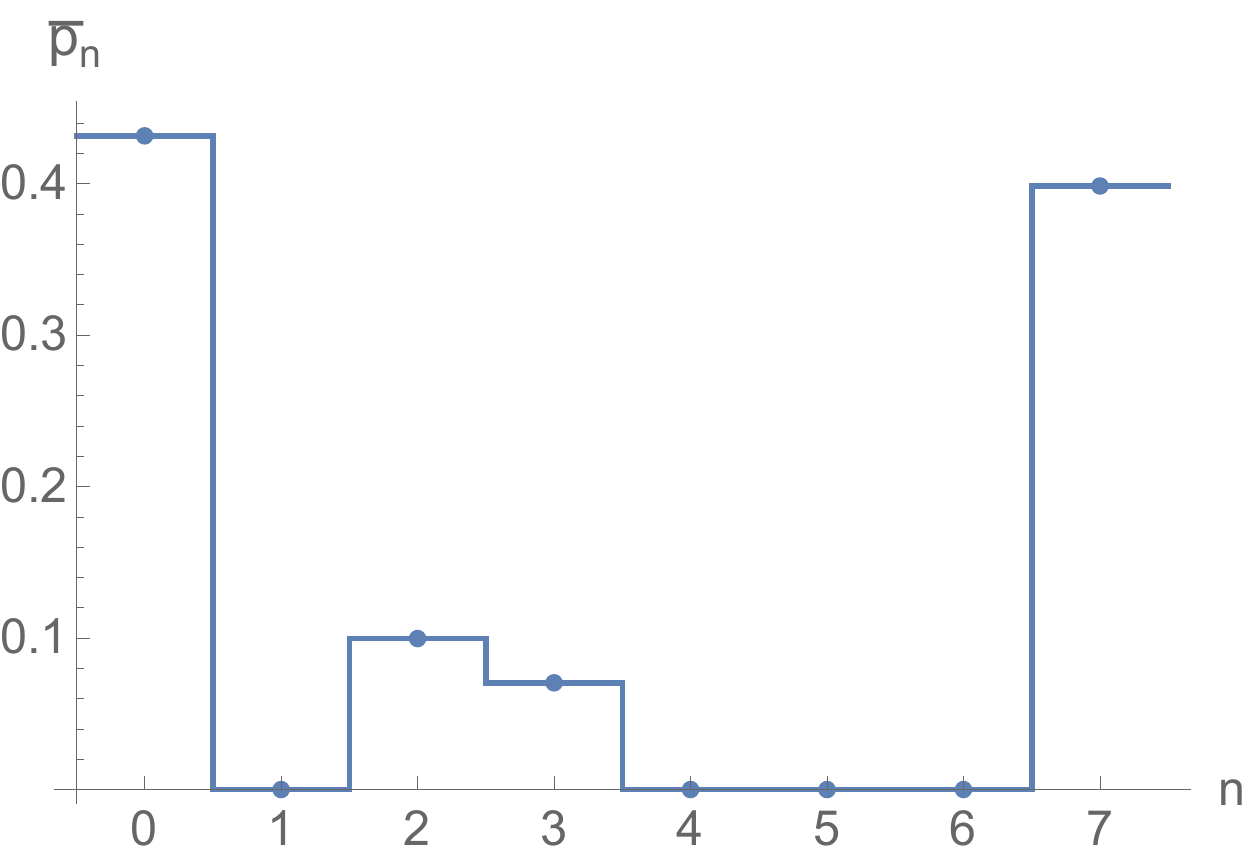}
  \caption{Optimal prior distribution for the encoding on the direct
    basis $B$ for a damping channel with
    hypergeometric decay ($M=5$ and $L=12$), in dimension $d=8$. Among
    the eight possible input states, just four ($n=0,2,3,7$) are used for the
    encoding. The corresponding detected capacity is given by
    $C_{DET}\simeq 1.074$ bits.}    
\end{figure}

Typically, for increasing values of damping the optimal prior
distribution for direct encoding shows holes of zero or negligible
probability, as depicted in Fig. 5 for the case $M=5$ and $L=12$, with
$d=8$.  This can be intuitively understood since in the presence of
strong damping it becomes more convenient to use a smaller alphabet of
well-spaced letters in order to achieve a better distinguishability at
the receiver.

\subsection{Negative hypergeometric channel}
We consider now a damping channel with decay structure characterized
by negative hypergeometric distributions, namely
\begin{eqnarray}
  Q(m|n)= \frac{ \binom{m +M -1}{m}\binom{L- M -m}{n - m}}{\binom{L}{n}}
  \;,
\end{eqnarray}
with positive integers $M$ and $L$ such that $n \leq L-M$ (here also
both $M$ and $L$ could vary for different values of $n$). This
distribution is related to the probability of $m$ successes until $M$
failures occur in drawing without replacement from finite samples of
$L$ elements. The mean and variance are given by \cite{kemp}
\begin{eqnarray}
  &&\mu = n \frac {M}{L-n+1} \;,\\ & &
 \sigma ^2 =  \mu  \left (1 -\frac \mu n \right ) \frac {L+1}{L-n+2}
  \;.
\end{eqnarray}
Notice that the variance is larger with respect to the
binomial distribution.
In the limit  $M,L \rightarrow \infty$ with $M/L = 1-\gamma $ one
recovers the binomial distribution.
\par For this class of channels we generally find that the detected
capacity is achieved by the Fourier basis $\tilde B$.
The results for $d=8$ and $L=32$ vs $M$ are reported in Fig. 6.

\begin{figure}[htb]
  \includegraphics[scale=0.46]{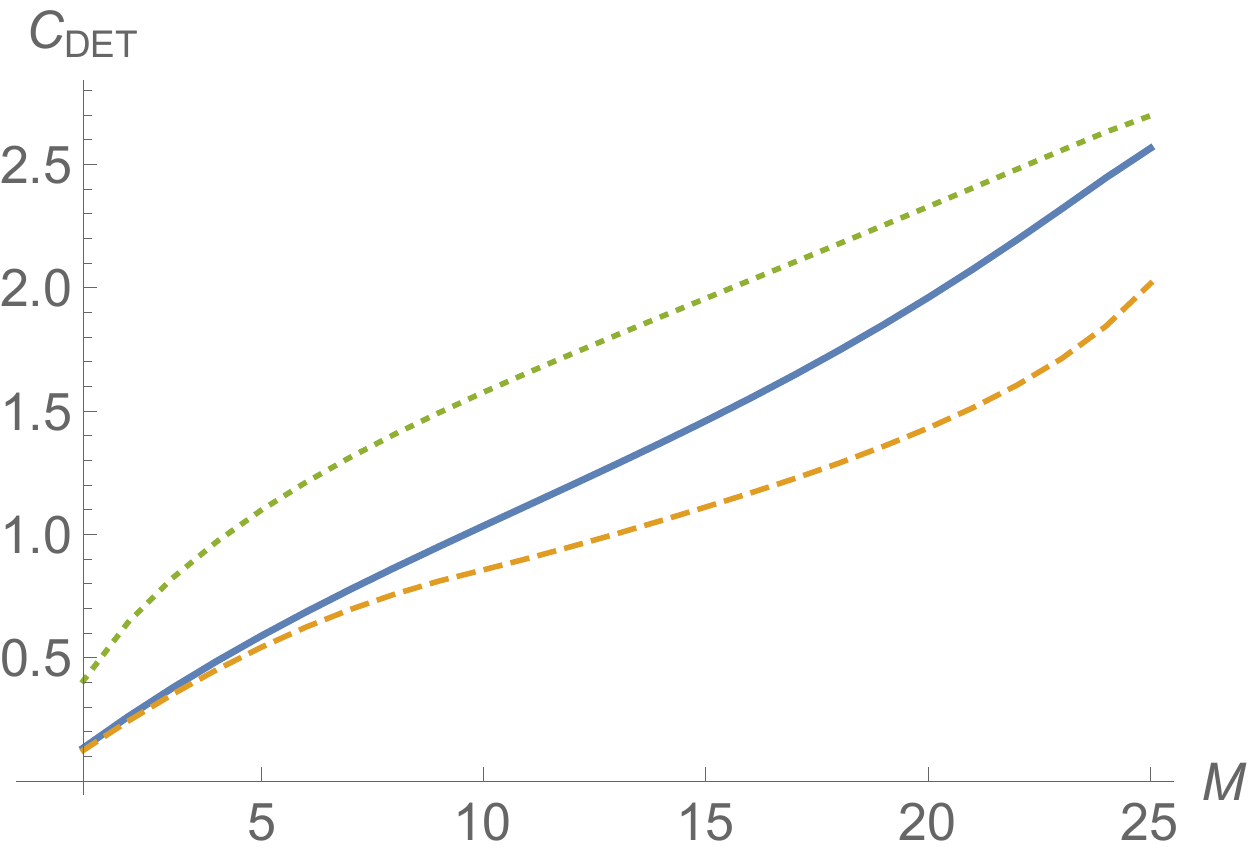}
  \caption{Detected classical capacity $C_{DET}$ for a damping channel
    with negative-hypergeometric decay for dimension $d=8$ and
    parameter $L=32$ vs $M$. The detected capacity is achieved by
    the Fourier Basis $\tilde B$ (solid line), which outperforms the direct basis
    $B$ (dashed line). The dotted line represents the theoretical lower bound given
  by the Holevo quantity $\chi _{\tilde B}$ of Eq. (\ref{holtil}).}
\end{figure}

\subsection{Beta-binomial channel}
We consider a damping channel with decay probabilities given by 
\begin{eqnarray}
  Q(m|n)= \binom{n}{m} \frac{B(m+\alpha , n-m +
    \beta )}{B(\alpha ,\beta )}
  \;,
\end{eqnarray}
where $\alpha , \beta >0$, and $B(\alpha ,\beta )=\Gamma (\alpha
)\Gamma (\beta )/\Gamma (\alpha + \beta )$ denotes the beta function. 
This family of distributions arises in binomial trials with success
probability that is not known, but distributed according to the beta
function. We remind that this distribution can be bimodal (U-shaped),
i.e. it can present two peaks when both $\alpha $ and $\beta $ are smaller
than 1.  The mean and variance are given by \cite{betab}
\begin{eqnarray}
  &&
  \mu = n \xi\;,\\
  &&\sigma ^2 =n \xi (1-\xi)\frac {\alpha +\beta +n}{\alpha +\beta +1}\,,
\end{eqnarray}
with $\xi =\frac {\alpha }{\alpha +\beta }$. We have then
overdispersion with respect to the binomial distribution with $\xi
=1-\gamma $. This binomial is recovered for $\alpha, \beta \rightarrow
\infty$ with $\xi =1-\gamma $. 
\par In Figs. 7 and 8 we plot the results of the detected capacity
$C_{DET}$ for dimension $d=8$
as a function of $\alpha $ and $\beta$. We notice that $C_{DET}$ is achieved 
by the Fourier basis $\tilde B$ (Fig. 7), except for a tiny region
corresponding to very small values of $\alpha $ and $\beta $ 
(Fig. 8).
\begin{figure}[htb]
  \includegraphics[scale=0.5]{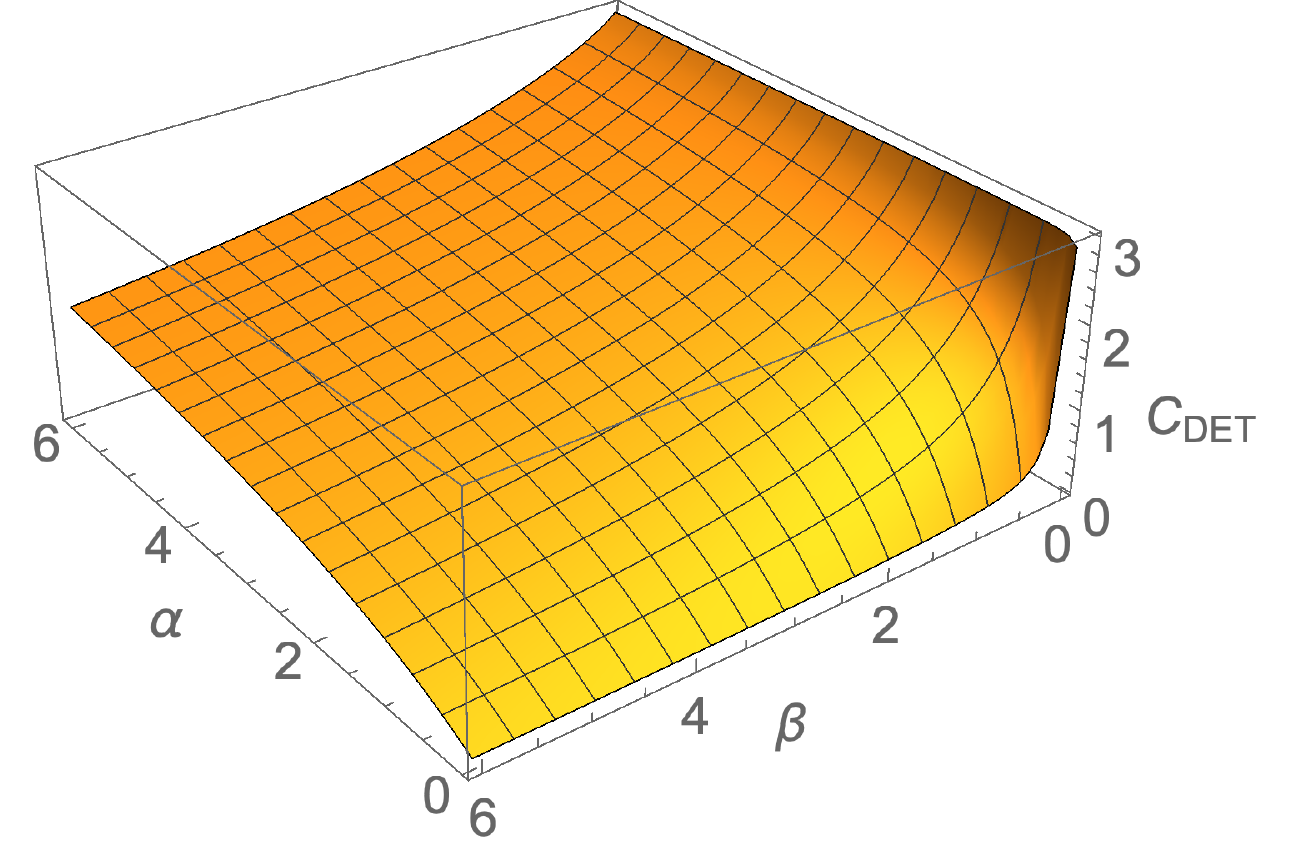}
  \caption{Detected classical capacity $C_{DET}$ for a beta-binomial
    decay channel with $d=8$ vs parameters $\alpha $ and $\beta $. In
    the present region the
    bound is provided by the Fourier encoding $\tilde B$.}
%
  \includegraphics[scale=0.46]{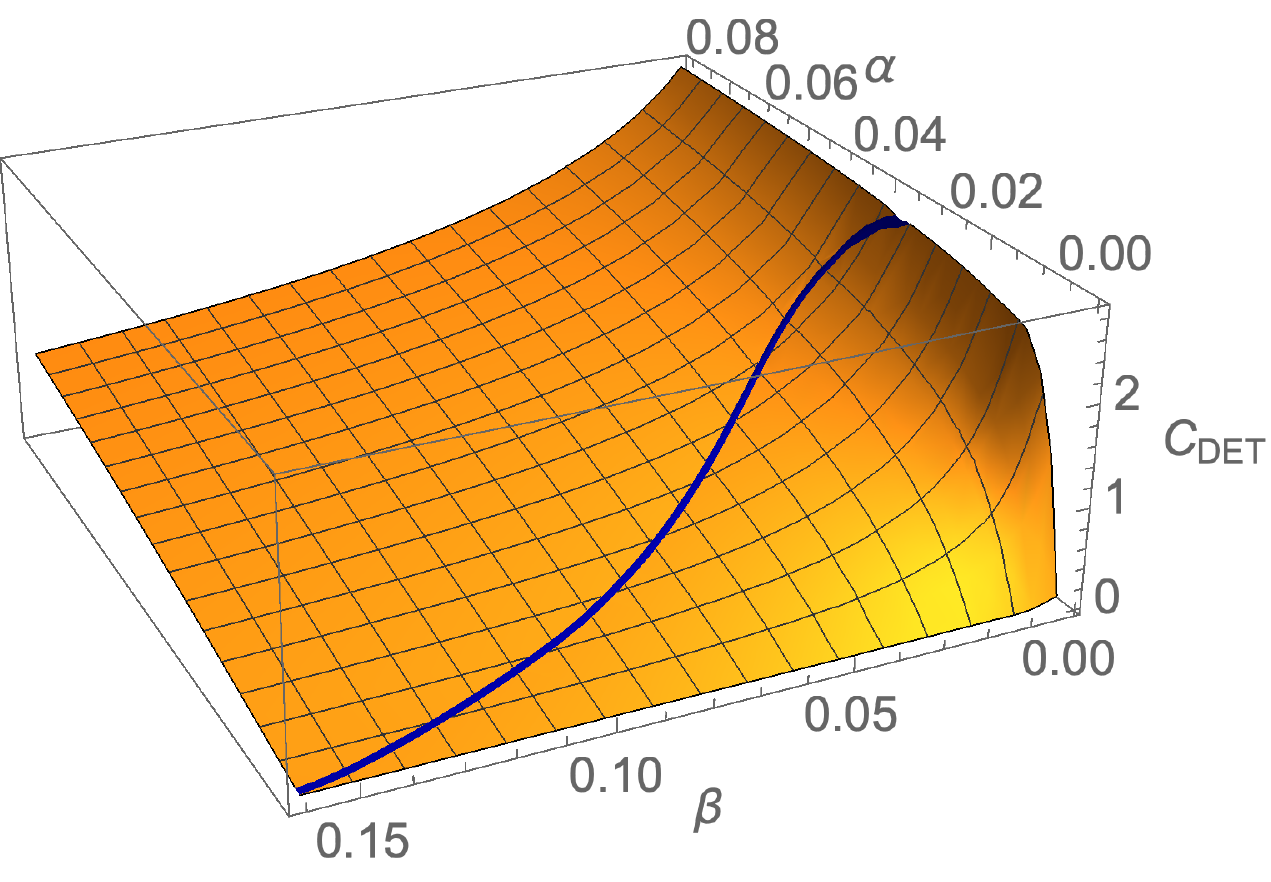}
  \caption{Detected classical capacity $C_{DET}$ for a beta-binomial
    decay channel with $d=8$ vs parameters $\alpha $ and $\beta $. In
    the region below the depicted line the bound is provided by the
    direct encoding $B$.}
\end{figure}
\par \noindent In Fig. 9 we also report the rescaled difference
$\Delta $ between the Holevo quantity $\chi _{\tilde B}$ and
$C_{DET}$.
\begin{figure}[htb]
  \includegraphics[scale=0.52]{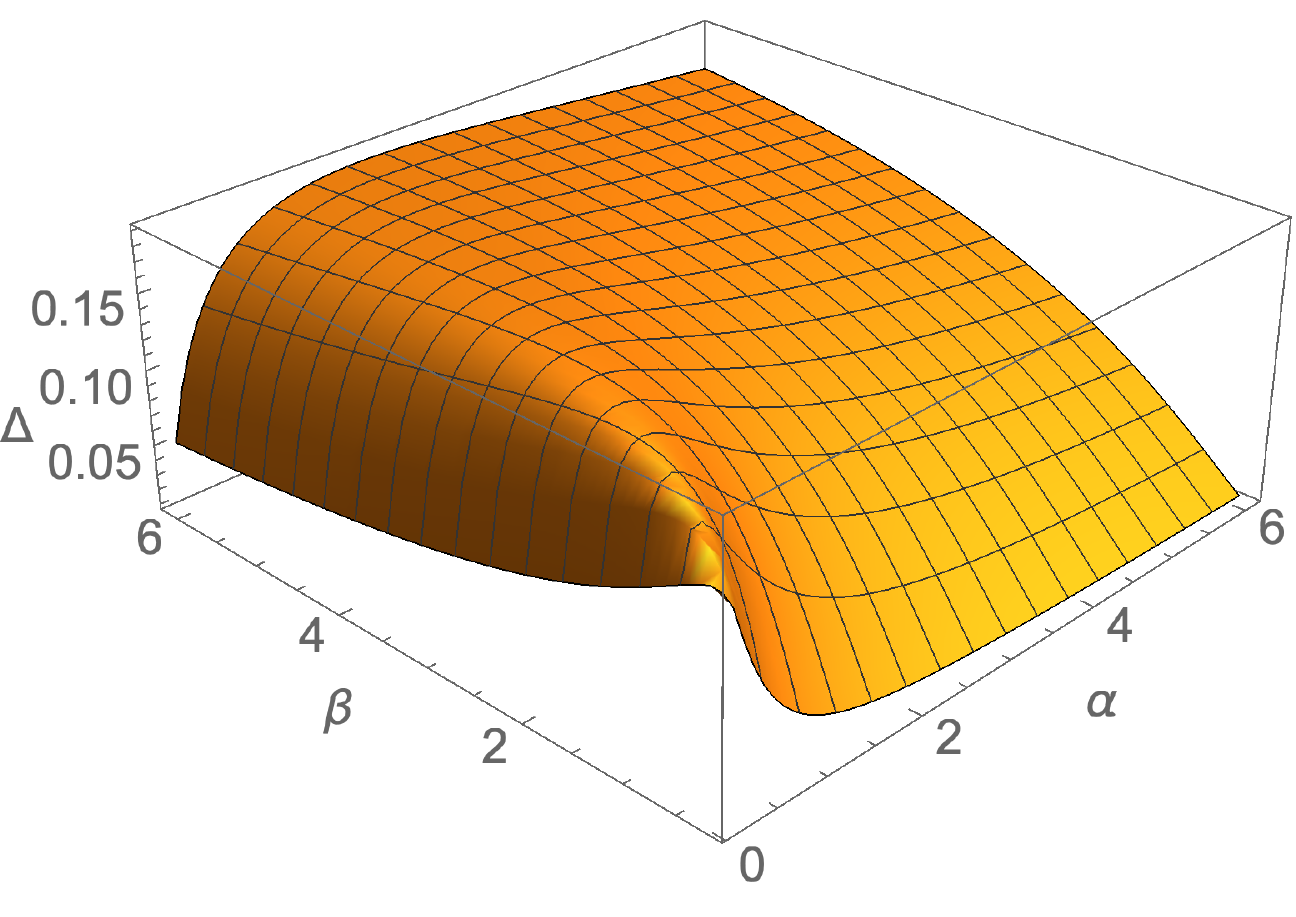}
  \caption{Rescaled difference $\Delta $ between the Holevo quantity
    $\chi _{\tilde B}$ and $C_{DET}$ represented in Fig. 7.}
\end{figure} 

\subsection{Geometric damping}
We consider a channel where the decaying conditional probabilities are
given by   
\begin{eqnarray}
  Q(m|n)=\frac {1- \gamma _n}{1- \gamma _n^{n+1}} \gamma _n ^{n-m} 
  \;,
\end{eqnarray}
with $\gamma _n \geq 0$.

\begin{figure}[htb]
  \includegraphics[scale=0.5]{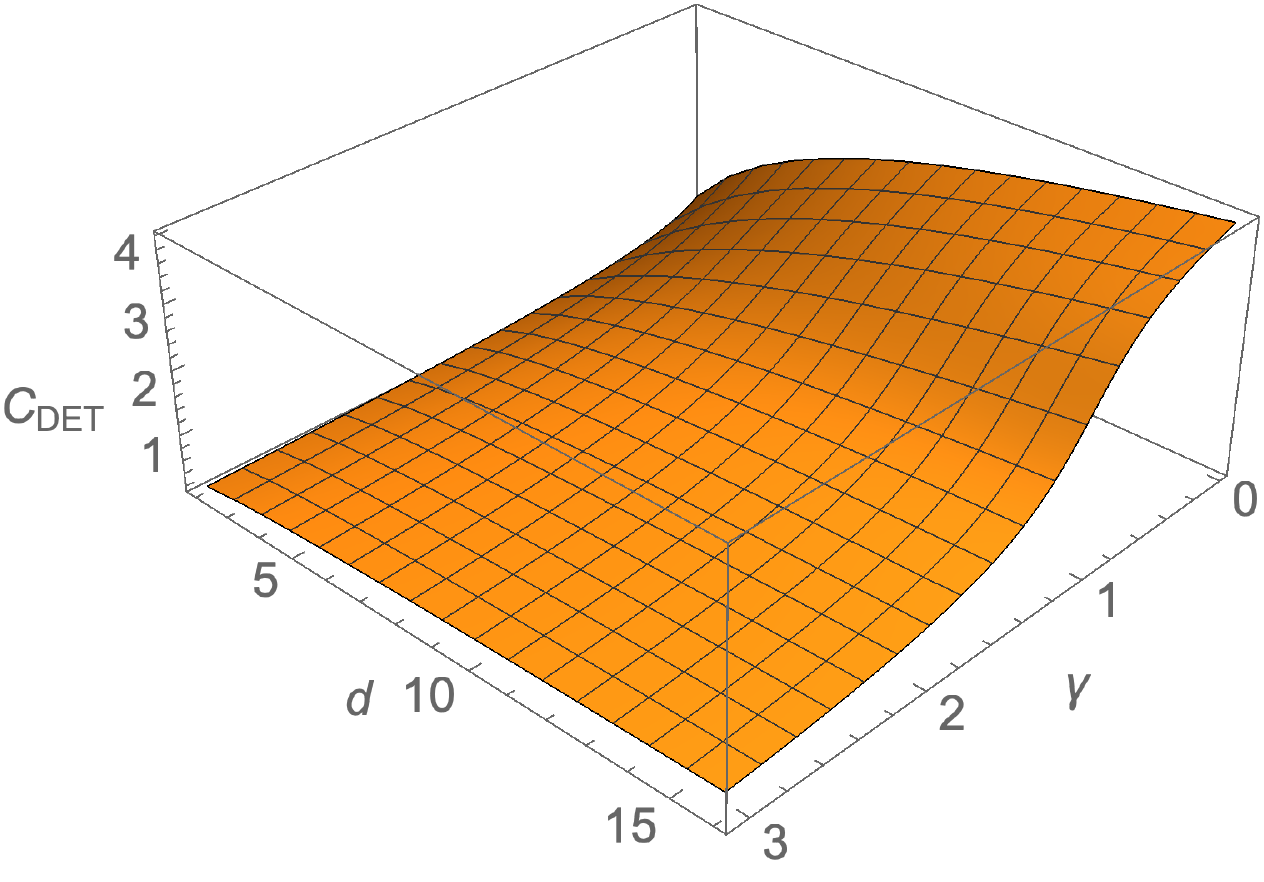}
  \caption{Detected classical capacity $C_{DET}$ for geometric damping
    channel vs dimension $d$ and decay parameters $\gamma _n =\gamma $, achieved
    by the Fourier basis $\tilde B $.}
%
  \includegraphics[scale=0.46]{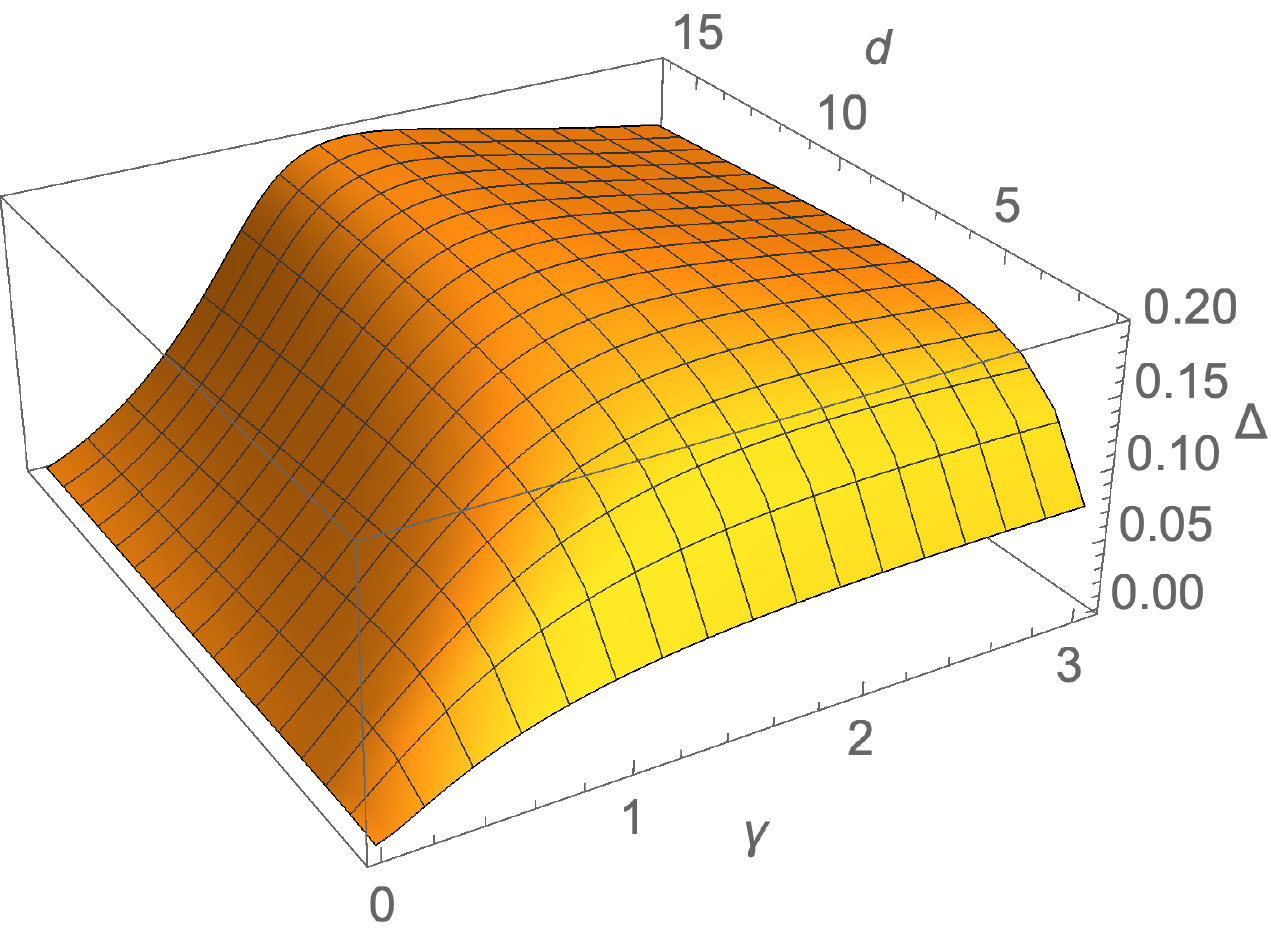}
  \caption{Rescaled difference $\Delta $ between the theoretical Holevo
    quantity $\chi _{\tilde B}$  and the detected classical capacity
    $C_{DET}$ for a geometric damping channel 
    vs dimension $d$ and damping $\gamma_n=\gamma $.}
\end{figure}
\par \noindent The results in the simplest case of $\gamma _n
=\gamma $ for all values of $n$ are depicted in Fig. 10, where the detected
capacity is always achieved by the Fourier basis $\tilde B$, for all
values of $\gamma $ and for any dimension $d$. In Fig. 11 we report
the rescaled difference with respect to the Holevo quantity $\chi
_{\tilde B}$.

\subsection{Constant ratio for adjacent levels}
We consider here a damping channel with constant ratio between the
decay probabilities pertaining to adjacent
levels, namely we study the case 
\begin{eqnarray}
\!\!\!\!\!\!\!
  Q(m|n)=\gamma_n ^{n-m}(1- \delta _{m,n})+ \frac{1-2 \gamma _n +
    \gamma _n^{n+1}}{1- \gamma _n}
  \delta _{m,n}
  \,,
\end{eqnarray}
with suitable positive values for $\gamma _n $. The result for $\gamma
_n=\gamma $ for all values of $n$ is reported in Fig. 12 for values of the
dimension $d=2,3,4$, and $5$ \cite{nota2}. We notice that, except for
the qutrit case $d=3$ with strong decay, the Fourier basis provides a
better lower bound to the channel capacity.
\begin{figure}[htb]
  \includegraphics[scale=0.5]{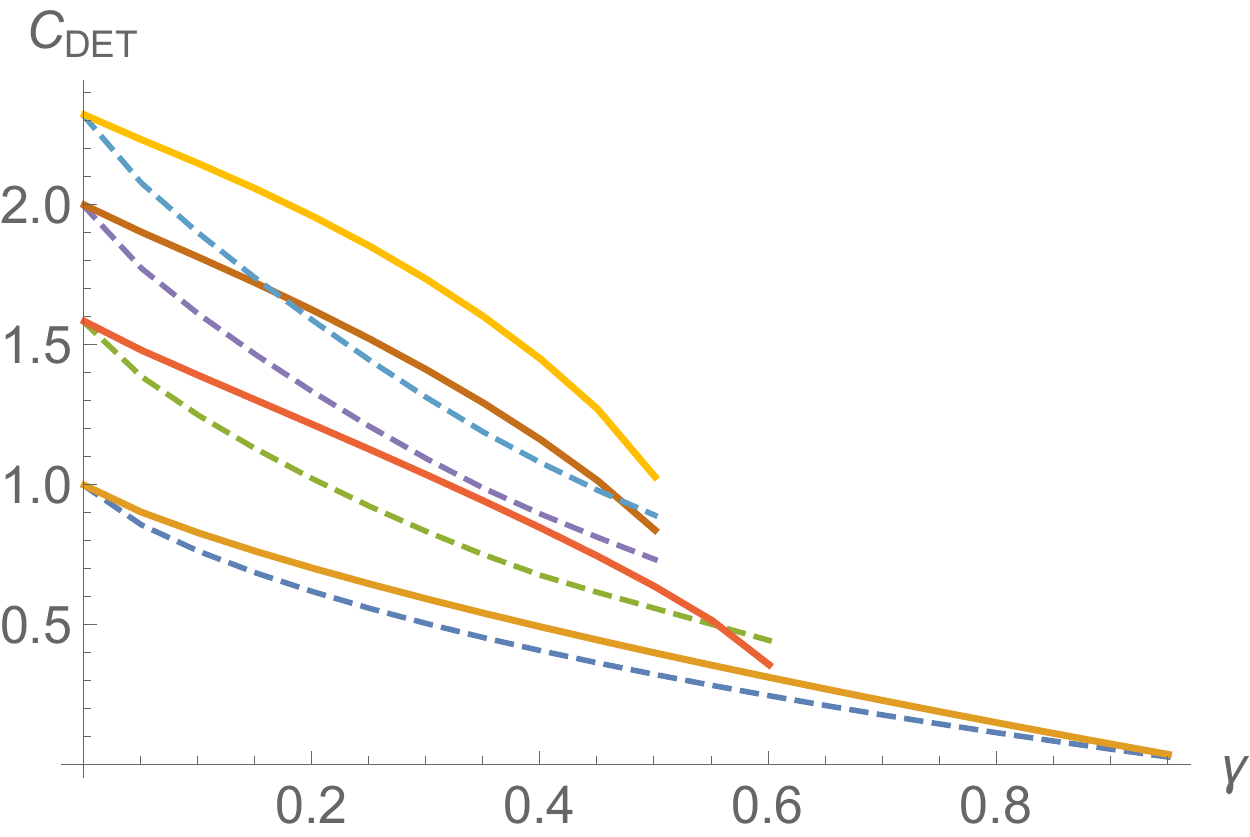}
  \caption{Detected classical capacity $C_{DET}$ for a constant-ratio
    decay channel vs allowed values of damping parameter $\gamma _n =\gamma
    $,  for
    $d=2,3,4,5$ (from bottom to top).  The solid (dashed) lines are referred to the
    Fourier $\tilde B$ (direct $B$) basis.}
  \end{figure} 

\subsection{Two-jump limited damping}
The following is an example of a damping channel where 
each level decays at most by two jumps: 
\begin{eqnarray}
&& Q(0|0)=1 \,,\nonumber \\& &
Q(m|1)=\frac{1}{1+\gamma _1}(\delta _{m,1}+ \gamma _1 \delta
_{m,0}) \,,\nonumber \\& &
Q(m|n)=\frac{1}{1+ \gamma _1 +\gamma _2}
(\delta _{m,n}+ \gamma _1 \delta
_{m-1,n} + \gamma _2 \delta _{m-2,n}) \nonumber \\& &
{\hbox{for }} 2 \leq n \leq d-1
\;,
\end{eqnarray}
with $\gamma _1,\gamma_2 \geq 0$.
The results of the detected capacity for dimension $d=8$ are reported
in Fig. 13. We observe a transition from the Fourier to the direct basis in
achieving the optimal detection for sufficiently large values of
$\gamma _1$ and $\gamma _2$.  

\begin{figure}[htb]
  \includegraphics[scale=0.44]{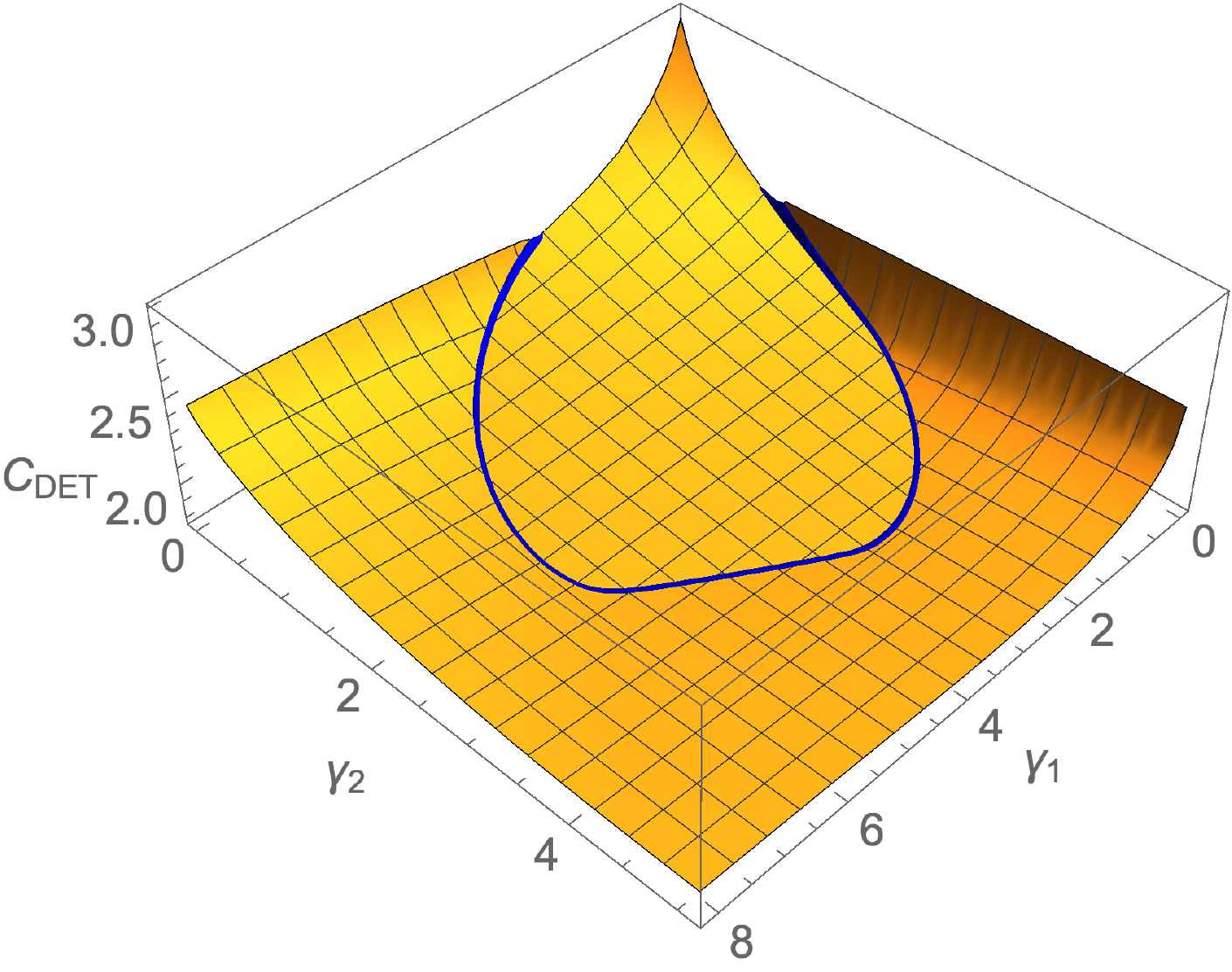}
  \caption{Detected classical capacity $C_{DET}$ for a two-jump
    limited decay channel with $d=8$ vs parameters $\gamma _1$ and $\gamma
    _2$. Inside (outside) the enclosed region the bound is achieved by the
    Fourier basis $\tilde B $ (direct basis $B$).}
\end{figure} 

\subsection{$\Lambda $-channels}
In this kind of damping channels only the uppermost level interacts
with each lower-lying level. Clearly, many variants are possible, and
we consider the following case
\begin{eqnarray}
&&   Q(m,d-1)=
  \frac {1- \gamma }{1- \gamma ^{d}} \gamma ^{d-1-m} 
  \;,\nonumber \\& &
  Q(m|n)= \delta _{m,n} \quad {\hbox{ for }} 0 \leq n<d-1\;,
\end{eqnarray}
with $\gamma \geq 0$. Indeed, this is a particular form of geometric
channel, where also the ratio of the transition
probabilities pertaining to adjacent levels is constant.
\begin{figure}[htb]
  \includegraphics[scale=0.46]{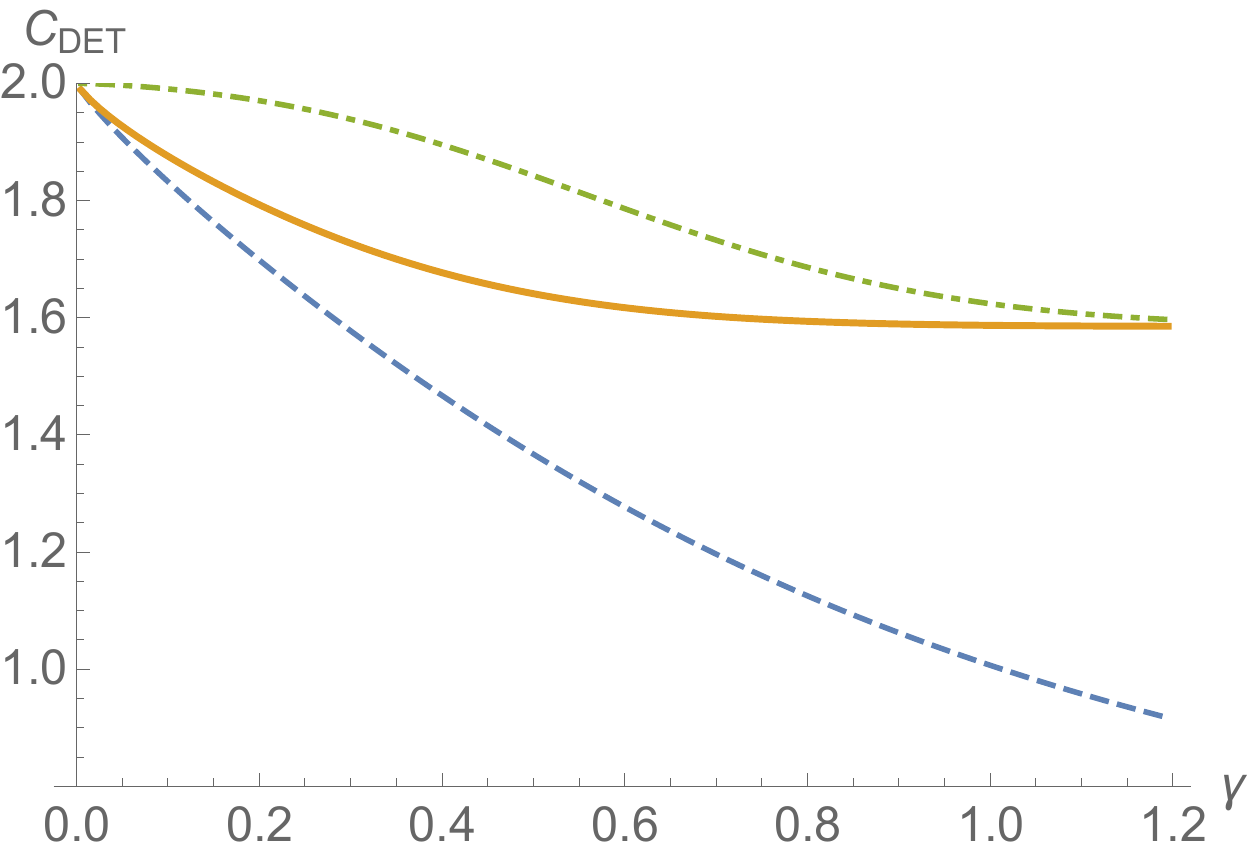}
  \caption{Detected classical capacity $C_{DET}$ for a
    $\Lambda $-channel vs damping parameter $\gamma $ for
    $d=4$ (solid line, achieved by the direct basis $B$). The
    looser bound in dashed line corresponds to the Fourier basis
    $\tilde B$. Since three levels are noise-free $C_{DET} > \log _2 3
    \simeq 1.585$ bits. 
    The Shannon entropy of the optimized prior probability $\{
    \bar p_{n}\}$ is depicted in dot-dashed line.} 
\end{figure} 
\par The solution for $d=4$ is depicted in Fig. 14. We notice that the detected capacity is
achieved by the direct basis $B$, for all values of $\gamma $. Interestingly, except for the qubit case $d=2$ (equivalent to the customary qubit damping channel),
 we have numerical evidence that
the direct basis always provides a better lower bound than the Fourier
basis for any $\gamma $ and $d$. 

\subsection{$V$-channels}
In this last example the lowest level is linked to a succession of higher-lying levels,
hence  
\begin{eqnarray}
  Q(m|n)=
  (1- \gamma _n) \delta _{n,n} + \gamma _n \delta _{n,0}
  \;,\
  \end{eqnarray}
with $\gamma _n \in [0,1]$. We considered the simplest case where
$\gamma _n=\gamma $ for all values of $n$, and the detected capacity
is plotted in Fig. 15 for values of the dimension $d=2,3,4,$ and
$8$. We notice that for $d=2$ and $3$ the values of the detected
capacity of Ref. \cite{ms19} are recovered. For increasing dimension
$d$ we observe a transition: except for the qubit case, where for any
$\gamma $ the best basis is the Fourier $\tilde B$, for $d>2$ the
direct basis $B$ rapidly outperforms $\tilde B$ for increasing values
of $\gamma $.

\begin{figure}[htb]
  \includegraphics[scale=0.5]{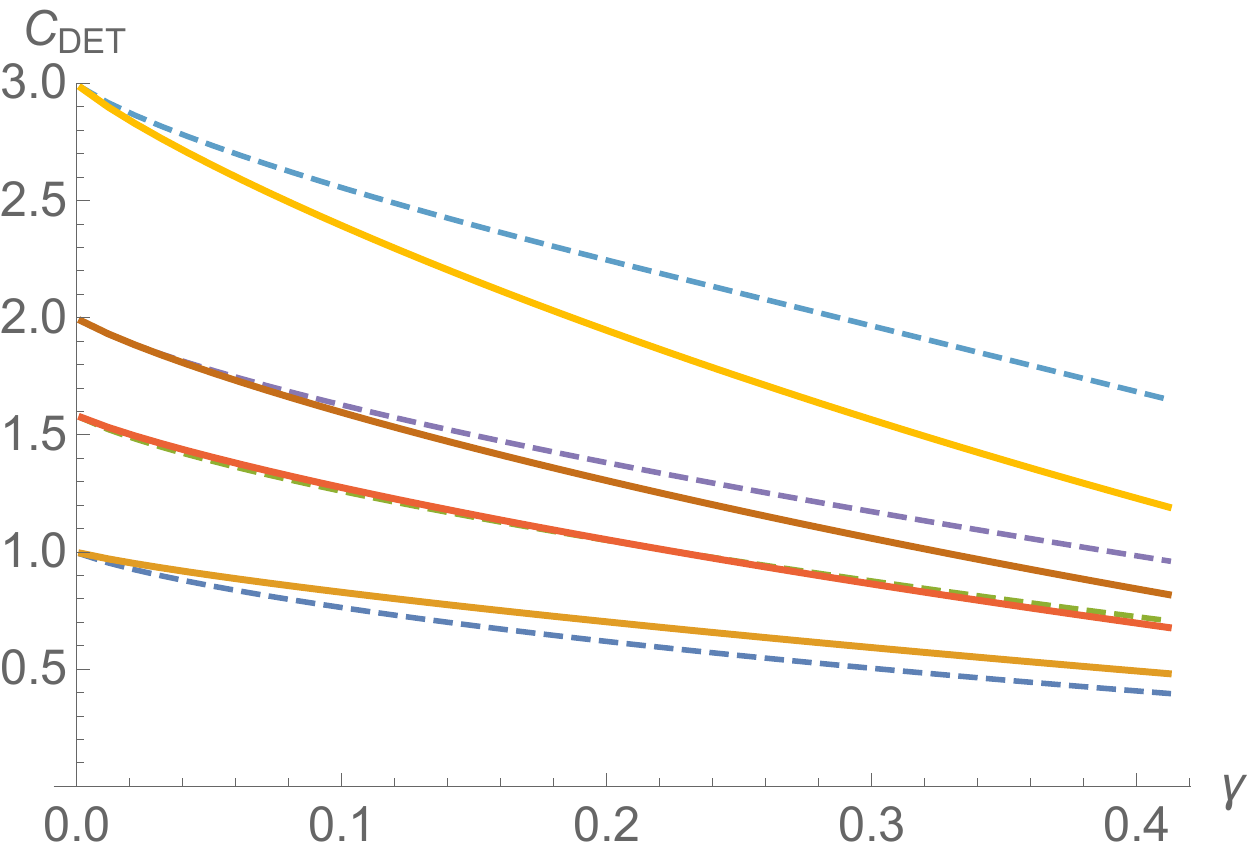}
  \caption{Detected classical capacity $C_{DET}$ for a
    $V$-channel vs damping parameter $\gamma $ for
    $d=2,3,4,8$ (from bottom to top).  The solid (dashed) lines are refereed to the
    Fourier $\tilde B$ (direct $B$) basis.}
    \end{figure}

\section{Conclusions}
We have applied a recently proposed general method \cite{ms19} to
detect lower bounds to the classical capacity of quantum communication
channels for general damping channels in dimension $d>2$. A number of
illustrative examples has been considered in the simplest scenario of
just two testing measurement settings, namely the direct coding on the
computational basis and on a Fourier basis. When the Fourier basis
$\tilde B$ outperforms the computational basis $B$, this gives an
indication that in such cases the accessible information for a single
use of the channel restricted to orthogonal input states and
projective output measurements can be improved by coding on
non-classical states with respect to the classical coding.  As a rule
of thumb, we observe that the Fourier basis provides a better lower
bound to the classical capacity as long as the variances of the
conditional probabilities $Q(m|n)$ pertaining to the direct coding are
sufficiently large. The present application to high-dimensional
channels strongly supports the use of our method especially when quantum complete
process tomography is unavailable or highly demanding, since, as we
have shown, remarkable results can be obtained by employing just 
two measurement settings. In general, by increasing the number of allowed testing
measurements, tighter bounds may be obtained. The method we employed 
is developed for unknown quantum channels.
Clearly, when some prior
knowledge about the structure of the channel is available, this could
be taken into account in the choice of selecting a limited number of
suited measurement settings.

\begin{acknowledgments}
C.M. acknowledges support by the European Quantera project QuICHE.
\end{acknowledgments}


\begin{thebibliography}{99}
\bibitem{nota}A quantum channel corresponds to a discrete
state change. Physically, it can be thought of as the integrated solution
of an open system dynamics (e.g., a Markovian master equation).
Depending on the model, in general the ``suitable limit'' in the first
case corresponds to a small-loss/short-time evolution scenario in the
interaction picture. The second case
corresponds to the Schroedinger picture, where in the limit only the
free Hamiltonian dynamics holds.
\bibitem{nota2}The curves cut off in different places for different
  values of $d$, because the allowed values of the damping parameter
  $\gamma $ depend on the dimension. In fact, Eq. (37) must provide proper normalized
conditional probabilities, and this forces different algebraic conditions on
$\gamma $ for different $d$.
\end{thebibliography}
\end{document}